\documentclass[11pt,a4paper]{article}
\usepackage[utf8]{inputenc}
\usepackage[english]{babel}
\usepackage{amsmath}
\usepackage{amsfonts}
\usepackage{amssymb}
\usepackage[pdftex]{graphicx}
\usepackage[font=small,labelfont=bf]{caption}
\usepackage{epstopdf}
\usepackage[T1]{fontenc}
\usepackage{etoolbox}
\usepackage{cleveref}
\usepackage{multirow}
\usepackage{amsbsy}
\usepackage[top=1.5in, bottom=1.5in, left=0.8in, right=0.8in]{geometry}
\usepackage{color}
\usepackage{lineno}
\usepackage{tikz}
\usetikzlibrary{patterns}
\usepackage{cleveref}
\usepackage{subcaption}
\usepackage{url}
\usepackage[bottom]{footmisc}
\usepackage{ctable} 
\usepackage{enumitem}
\usepackage{wrapfig}

\allowdisplaybreaks

\makeatletter
\renewcommand\tableofcontents{%
    \@starttoc{toc}%
}
\makeatother

\allowdisplaybreaks

\DeclareUnicodeCharacter{00A0}{~}
\DeclareUnicodeCharacter{2212}{~}

\begin{document}

\newcommand{\ipnp}{\ensuremath{_{i+1}^{n+1}}}
\newcommand{\imnp}{\ensuremath{_{i-1}^{n+1}}}
\newcommand{\inp}{\ensuremath{_{i}^{n+1}}}

\newcommand{\ipn}{\ensuremath{_{i+1}^{n}}}
\newcommand{\imn}{\ensuremath{_{i-1}^{n}}}
\newcommand{\inn}{\ensuremath{_{i}^{n}}}

\newcommand{\np}{\ensuremath{^{n+1}}}
\newcommand{\n}{\ensuremath{^{n}}}

\newcommand{\subij}{\ensuremath{_{i,\,j}}}
\newcommand{\subimj}{\ensuremath{_{i-1,\,j}}}
\newcommand{\subipj}{\ensuremath{_{i+1,\,j}}}
\newcommand{\subijm}{\ensuremath{_{i,\,j-1}}}
\newcommand{\subijp}{\ensuremath{_{i,\,j+1}}}

\newcommand{\erf}{\ensuremath{\text{erf}}}

\newcommand{\etal}{\emph{et al.}$\;$}

\newcommand{\ud}{\mathop{}\!\mathrm{d}} 

\newcommand{\red}{\textcolor{red}}
\newcommand{\blue}{\textcolor{blue}}

\newcommand{\highlight}[1]{%
  \colorbox{gray!25}{$\displaystyle#1$}}
  
\newcommand\T{\rule{0pt}{5ex}}

\providecommand{\keywords}[1]{\hspace{2.5mm} \textbf{Keywords} #1}

\newcommand*\samethanks[1][\value{footnote}]{\footnotemark[#1]}

%\makeatletter
%\patchcmd{\maketitle}{\@fnsymbol}{\@alph}{}{}  % Footnote numbers from symbols to small letters
%\makeatother

\title{Cellular automaton model for substitutional binary diffusion in solids}

\author{H. Ribera \thanks{Centre de Recerca Matemàtica, Campus de Bellaterra, Edifici C, 08193 Bellaterra, Barcelona, Spain.}$\,\,^{,}\,$\thanks{Department de Matemàtica Aplicada, Universitat Politècnica de Catalunya, Barcelona, Spain.} \and B. Wetton \thanks{Mathematics Department, University of British Columbia, Vancouver BC, Canada V6T 1Z2.} \and T.~G. Myers \footnotemark[1]$\,\,^{,}\,$\footnotemark[2]}

\date{\today}

\maketitle

%%%%%%%%%%%%%%%%%%%%

\begin{abstract}

In this paper we use the cellular automaton (CA) approach to model one-dimensional binary diffusion in solids. Employing a very simple state change rule we define an asynchronous CA model and take its continuum limit to obtain the governing equations of the problem. We show that in the limit where the number of cells tends to infinity the CA model approaches a continuous model derived in previous work \cite{Ribera2017}. Thus, showing that the CA approach provides a new, simple method to study and model binary diffusion.

\end{abstract}


%%%%%%%%%%%%%%%%%%%%

%\vspace{10mm} 
%
%\noindent\rule{\textwidth}{0.4pt}
%
%\vspace{1mm} 
%\renewcommand{\baselinestretch}{0.3}
%\tableofcontents
%\renewcommand{\baselinestretch}{1.2}
%\vspace{1mm} 
% 
%\noindent\rule{\textwidth}{0.4pt}
 
%%%%%%%%%%%%%%%%%%%%

\section{Introduction}

A cellular automaton (CA) model consists of an $n$-th dimensional space partitioned into a discrete subset of $n$-dimensional volumes, which are called cells and are defined in a discrete time. A finite list of possible states is defined for each cell, and each cell has one state. A local neighbourhood is defined for each cell at every time step. The state of a cell can be changed by a \emph{state change rule}, which is a rule that allows the computation of the new state for the cell, and is dependent on other cells in the local neighbourhood \cite{Janssens2010}. Typically, this rule is fixed, that is, it is the same rule for all cells. It does not change over time and it is applied to all cells simultaneously. However, this rule can be stochastic, which means that the new states are chosen according to some probability distribution. This rule can also be applied to each individual cell independently and so the new state of a cell affects the calculation of states in neighbouring cells. Chopard \etal \cite{Chopard1989} presented the first application of cellular automaton to model diffusion on lattices. Subsequently many other CA approaches were applied to reaction-diffusion problems, see Boon \etal \cite{Boon1996} and Weimar \cite{Weimar1997}. Then, it seems, there is a pause in the literature on using cellular automata approaches to work with the diffusion equation. Moreover, this type of modelling does not seem to have been applied to work with coupled, nonlinear diffusion problems, which are the focus of this paper.

The \emph{Kirkendall effect} is the name given to the physical phenomenon whereby atomic diffusion occurs via a vacancy exchange mechanism instead of a substitutional or ring mechanism. Recently the Kirkendall effect has been used to create hollow nanostructures, which can be used in a variety of applications. In Aldinger \cite{Aldinger1974} they present the first use of the Kirkendall effect to create hollow structures. The hollow permits their use in transporting drugs and biomolecules and then releasing them in a controlled manner \cite{An2009}. Na \etal \cite{Na2009} present a review on nanostructrues and magnetic resonance imaging (MRI). These structures have also been proposed to enhance the rate capability and cycling stability in lithium-ion batteries \cite{Wang2012}. Hollow nanoparticles have also been reported to be good catalysts \cite{Kim2002,Li2006}. 

In an attempt to understand, and so better control the growth of hollow nanoparticles Ribera \etal \cite{Ribera2017} rigorously derived governing equations for the substitutional binary diffusion problem. Moreover, under sensible assumptions they reduce these governing equations in order to provide an analytically tractable problem. As a starting point they examine the one-dimensional problem of an insulated bar. In this paper we investigate the same problem but from the cellular automaton standpoint. This will help understand further the physical mechanisms behind the Kirkendall effect.

In the following section we present the model of Ribera \etal \cite{Ribera2017}, which will be the starting point of the cellular automaton model, presented in Section \ref{sec-CA}. In the limit where the number of cells is large we prove that the CA model reduces to a special case of binary diffusion, where one species diffuses much faster than the other. In the Results section we verify this by showing that at large time and with a fast diffuser the CA model coincides with the continuum model. Further, since the computer speed reduces with $N$ it is clear that the CA model is particularly useful at small scales, such as with nanostructures.
%We will extend the aforementioned model to its continuum limit. In the Results section we compare this limit to the solution of the continuum model in \cite{Ribera2017}, and show that both approaches lead to the same governing equations thus providing a powerful tool to model various diffusion processes, such as the one that leads to the creation of nanostructures.

%%%%%%%%%%%%%%%%%%%%%%%%%%%%%%%%%%%%%%%%%%%%%%%%%%%%%%%%%%%%%%%%%%%%%

\section{Continuum model for substitutional binary diffusion}
\label{sec-model-reduced}

Let us consider a binary crystalline solid composed of three species: atomic species A, atomic species B, and vacancies V. We label the fast diffuser as species A, and the slow one B. Since we are considering a perfect lattice, that means that the sum of all the fluxes is zero and then it is only necessary to work with the evolution of two species to fully define the problem. %The governing equations in terms of A and V are,
%\begin{align}
%\label{gov-eq-ficks-1}
%\frac{\partial X_A}{\partial t} &= \nabla \cdot \left( D_{AA}^V \nabla X_A \right) - \nabla \cdot \left( D_{AV} \nabla X_V \right),\\
%\label{gov-eq-ficks-2}
%\frac{\partial X_V}{\partial t} &= -\nabla \cdot \left( D_{VA} \nabla X_A \right) + \nabla \cdot \left( D_{VV} \nabla X_V \right),
%\end{align}
%where $X_i$ are the mole fractions corresponding to the $i$-th species, and $D_{AA}^V$, $D_{AV}$, $D_{VA}$, $D_{VV}$ are the diffusion coefficients. The definition of these diffusion coefficients can be found in equations \red{(??)-(??)} in \cite{Ribera2017} \red{[WAITING TO FINISH THAT PAPER OFF FIRST TO KNOW THE NUMBER OF THE EQUATIOS].} In the following we will study the one dimensional case under the assumption that A has a much higher diffusion rate than B.

%\subsection{One-dimensional case}
%\label{sec-model-reduced}

Now consider an insulated one-dimensional bar of length $2l$. At $t=0$ the side $x \in [-l,0]$ is made of material A (and a proportion of vacancies), and the side $x \in [0,l]$ is made of material B (and a proportion of vacancies). %See Figure \ref{figure-model-drawing} for a sketch of the situation.

For $t >0$ the diffusion of species is defined by 
\begin{align}
\label{gov-eq-1-mathmodel}
\frac{\partial X_A}{\partial t} &= \frac{\partial}{\partial x}\left(D_{AA}^V \frac{\partial X_A}{\partial x} \right) - \frac{\partial}{\partial x}\left(D_{AV} \frac{\partial X_V}{\partial x} \right),\\
\label{gov-eq-2-mathmodel}
\frac{\partial X_V}{\partial t} &= - \frac{\partial}{\partial x}\left(D_{VA} \frac{\partial X_A}{\partial x} \right) + \frac{\partial}{\partial x}\left(D_{VV} \frac{\partial X_V}{\partial x} \right),
\end{align}
where $X_i$ are the mole fractions corresponding to the $i$-th species and the diffusion coefficients $D_{AA}^V$, $D_{AV}$, $D_{VA}$ and $D_{VV}$ vary nonlinearly with $X_i$ \cite{Ribera2017}. In the limit where A diffuses much faster than B,
\begin{equation}
\label{reduced-diffusion}
\begin{split}
\;\;\;\;\;\;\;\: \hat{D}_{AA}^V &\sim \Gamma X_V, \;\;\;\;\;\;\;\;\;\;\;\;\;\;\;\;\;\;\;\;\;\;\;\:\! \hat{D}_{AV} \sim \Gamma X_A,\\
\hat{D}_{VA} &\sim \left( \Gamma - 1 \right) X_V, \;\;\;\;\;\;\;\;\;\;\;\;\;\, \hat{D}_{VV} \sim \left[\left(\Gamma - 1 \right)X_A + 1 \right].
\end{split}
\end{equation}
%Note, $X_B = 1 - X_A - X_V$. 
The boundary conditions are
\begin{equation}
\label{bc}
\frac{\partial X_A}{\partial x}\bigg|_{x=\pm l} = \frac{\partial X_V}{\partial x}\bigg|_{x=\pm l} = 0,
\end{equation}
and the initial conditions are
\begin{equation}
\begin{aligned}
\label{ic}
X_A(x,0) = \left\{
     \begin{array}{lcr}
      X_{A,\text{ini}} & \text{if} & -1 < \,\, x < 0,\\
      0 & \text{if} & 0 < \,\, x < 1,
     \end{array}
   \right. \;\;\;&\;\;\;
X_B(x,0) = \left\{
     \begin{array}{lcr}
      0 & \text{if} & -1 < \,\, x < 0,\\
      X_{B,\text{ini}} & \text{if} & 0 < \,\, x < 1,
     \end{array}
   \right. \\   
   \; \\
   X_V(x,0) &= X_{V,\text{ini}},   
\end{aligned}
\end{equation}
where $X_{A,\text{ini}}$, $X_{B,\text{ini}}$, and $X_{V,\text{ini}}$ denote the constant initial mole fractions of material A, B, and vacancies, respectively, and $X_{i,\text{ini}} = 1 - X_{V,\text{ini}}$, for $i=$ A, B. 

In the following section we will show that the CA model reduces to \eqref{gov-eq-1-mathmodel}, \eqref{gov-eq-2-mathmodel} in the appropriate limit.

%%%%%%%%%%%%%%%%%%%%%%%%%%%%%%%%%%%%%%%%%%%%%%%%%%%%%%%%%%%%%%%%%%%%%%

\section{Cellular automaton model}
\label{sec-CA}

In order to define the cellular automaton model for the one-dimensional problem discussed in \S \ref{sec-model-reduced} we are going to define a two-dimensional space of size $N\times N$ that is partitioned into two-dimensional $1\times 1$ cells. This would correspond to a two dimensional lattice of $N\times N$ atoms, in which each cell corresponds to one atomic site. Thus, the list of states for each cell in the CA model are ``A atom'', ``B atom'', and ``vacancy V''. The grid is considered to be periodic on the top/bottom edges. The model presented here will be asynchronous, that is, at each time step only one cell will be picked to apply the state change rule. Since physically atomic diffusion happens via vacancy exchange, it makes sense that the only cells in our CA grid that change state are those situated next to a vacancy cell and the vacancy cells themselves. For this reason, at each time step we only pick cells that represent vacancies to apply the change of state rule. Moreover, the choice of which vacancy cell is picked is done at random. We define the local neighbourhood of a cell as all the cells that surround it. Thus, each cell has eight neighbours, except the ones on the left and right columns on the grid, which can have five or three (corners) neighbours. We pick one of these neighbours at random and then apply the state change rule, which is defined as follows. If the neighbour cell picked is an A cell, we will proceed to exchange the states of the vacancy and A cell, and so A has moved. If the neighbour cell picked is a B cell, the probability of exchanging states with the vacancy cell is defined to be $1/\Gamma$. This will capture the physical feature in the model of B being $\Gamma$ times slower than A. Finally, if the neighbour cell picked is a vacancy, no change of states is applied. 

\begin{figure}[!htb]
\centering
\begin{tikzpicture}[scale=0.5, transform shape]
%\draw (-8.5,-8.5) -- (8.5,-8.5) -- (8.5,6.5) -- (-8.5,6.5) -- (-8.5,-8.5); 
\draw (-8.5,-5) -- (8.5,-5);
\draw (-8.5,-3) -- (8.5,-3);
\draw (-8.5,-1) -- (8.5,-1);
\draw (-8.5,1) -- (8.5,1);
%\draw (-8.5,3) -- (8.5,3);
\draw (-8.5,5) -- (8.5,5);
\draw (-5,6.5) -- (-5,-8.5);
\draw (-3,6.5) -- (-3,-8.5);
\draw (-1,6.5) -- (-1,-8.5);
\draw (1,6.5) -- (1,-8.5);
\draw (3,6.5) -- (3,-8.5);
\draw (5,6.5) -- (5,-8.5);

\draw[line width=0.75mm] (-5.5,3) -- (5.5,3);
\draw[dashed, line width=0.75mm] (-5.5,3) -- (-6.5,3);
\draw[dashed, line width=0.75mm] (5.5,3) -- (6.5,3);
\draw[line width=0.75mm] (6.5,3) -- (7,3);
\draw[line width=0.75mm] (7,3) -- (7,-5.5);
\draw[dashed, line width=0.75mm] (7,-5.5) -- (7,-6.5);
\draw[line width=0.75mm] (7,-6.5) -- (7,-7);
\draw[line width=0.75mm] (7,-7) -- (6.5,-7);
\draw[dashed,line width=0.75mm] (6.5,-7) -- (5.5,-7);
\draw[line width=0.75mm] (5.5,-7) -- (-5.5,-7);
\draw[dashed, line width=0.75mm] (-5.5,-7) -- (-6.5,-7);
\draw[line width=0.75mm] (-6.5,-7) -- (-7,-7);
\draw[line width=0.75mm] (-7,-6.5) -- (-7,-7);
\draw[dashed,line width=0.75mm] (-7,-6.5) -- (-7,-5.5);
\draw[line width=0.75mm] (-7,3) -- (-7,-5.5);
\draw[line width=0.75mm] (-7,3) -- (-6.5,3);
\draw[line width=0.75mm] (-7,3) -- (-7,5.5);
\draw[dashed, line width=0.75mm] (-7,5.5) -- (-7,6.5);
\draw[line width=0.75mm] (7,3) -- (7,5.5);
\draw[dashed, line width=0.75mm] (7,5.5) -- (7,6.5);
\draw[line width=0.75mm] (-7,-7) -- (-7,-7.5);
\draw[dashed, line width=0.75mm] (-7,-7.5) -- (-7,-8.5);
\draw[line width=0.75mm] (7,-7) -- (7,-7.5);
\draw[dashed, line width=0.75mm] (7,-7.5) -- (7,-8.5);
\draw[line width=0.75mm] (-7,3) -- (-7.5,3);
\draw[dashed, line width=0.75mm] (-7.5,3) -- (-8.5,3);
\draw[line width=0.75mm] (7,3) -- (7.5,3);
\draw[dashed, line width=0.75mm] (7.5,3) -- (8.5,3);
\draw[line width=0.75mm] (-7,-7) -- (-7.5,-7);
\draw[dashed, line width=0.75mm] (-7.5,-7) -- (-8.5,-7);
\draw[line width=0.75mm] (7,-7) -- (7.5,-7);
\draw[dashed, line width=0.75mm] (7.5,-7) -- (8.5,-7);

\node at (0, 2) {\Huge V};
\node at (-4.75, -6) {$(i,j)$ subgrid};
\node at (-5.1, 5.4) {$(i,j+1)$ subgrid};
\node at (-5.1, -8) {$(i,j-1)$ subgrid};
\node at (-9, 2) {$(i-1,j)$ subgrid};
\node at (9, 2) {$(i+1,j)$ subgrid};

\node at (0, -9) {$\leftarrow \sqrt{N} \rightarrow$};
\node at (9, -1.5) {$\uparrow$};
\node at (9, -2) {$\sqrt{N}$};
\node at (9, -2.5) {$\downarrow$};

\draw [line width=0.75mm, red] (-3,5) -- (3,5);
\draw [line width=0.75mm, red] (-3,-1) -- (3,-1);
\draw [line width=0.75mm, red] (-3,5) -- (-3,-1);
\draw [line width=0.75mm, red] (3,5) -- (3,-1);

\end{tikzpicture}
\caption{Sketch of the subgrid set-up. In red the local neighbourhood of a V cell is shown.}
\label{figure-subgrid}
\end{figure}
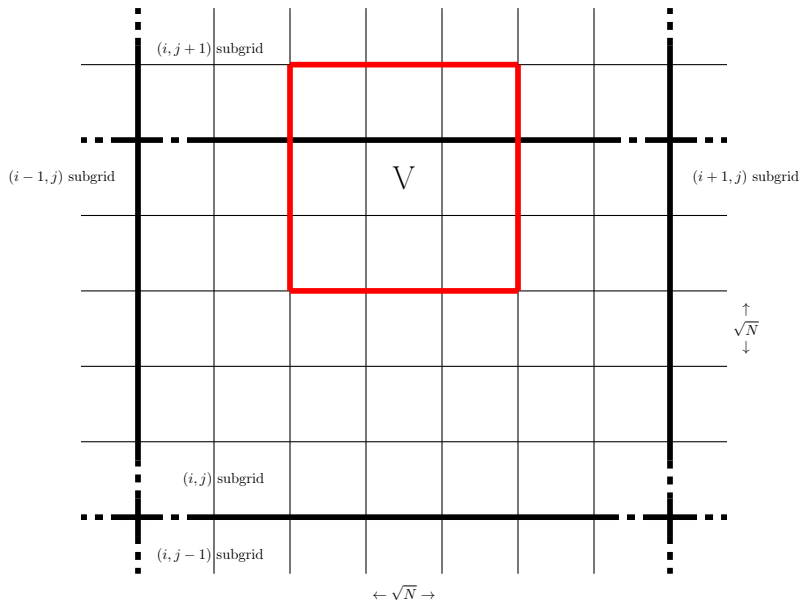

Our interest now is to find the continuum limit of the asynchronous cellular automaton model we have described. Let us define the fraction of V cells, A cells, and B cells in the whole grid as
\begin{equation}
\bar{V} = \frac{N_V}{N^2}, \qquad \bar{A} = \frac{N_A}{N^2}, \qquad \bar{B} = \frac{N_B}{N^2},
\end{equation}
where $N_i$ is the number of cells of state $i$. Let us pick a square subgrid of size $\sqrt{N} \times \sqrt{N}$, and name it the $(i,j)$ subgrid (see Figure \ref{figure-subgrid}). Inside it, we define the following three functions,
\begin{itemize}
\setlength\itemsep{0em}
\item $\bar{V}\subij$, the fraction of V cells in the $(i,j)$ subgrid;
\item $\bar{A}\subij$, the fraction of A cells in the $(i,j)$ subgrid;
\item $\bar{B}\subij$, the fraction of B cells in the $(i,j)$ subgrid.
\end{itemize}
All three functions above are dependent on space and time. Note that the choice of $M=\sqrt{N}$ of the subgrid is arbitrary. We only need $\lim_{N\rightarrow \infty} M/N = 0$ in a suitable manner.

We wish to study the evolution in one time step of the fraction number of V cells, A cells and B cells inside the $(i,j)$ subgrid. Let $\bar{A}\subij^n$, $\bar{V}\subij^n$, $\bar{B}\subij^n$ be the fraction number of A cells, vacancies and B cells, respectively, in the $(i,j)$ subgrid at time step $n$. The aim is to compute $\bar{A}\subij\np$, $\bar{V}\subij\np$ and $\bar{B}\subij\np$. In the next two sections we will discuss the change of A cells and V cells in one time step inside the  $(i,j)$ subgrid, respectively. We will omit the case of B cells since by conservation it can be found from A and V. %it can be found as a result of the first two as it was the case in the substitutional diffusion model presented in \S \ref{sec-model-reduced}. 

\subsection{A cells}

There are two factors that can affect the amount of A cells in the $(i,j)$ subgrid in one time step: either a vacancy of the subgrid is able to exchange places with an A cell of a neighbouring subgrid (that adds an A cell), or a vacancy from a neighbouring subgrid is able to exchange places with an A cell in the $(i,j)$ subgrid (that removes an A cell). Thus, to add one A cell it is necessary that at the $n$-th time step
\begin{enumerate}[label={\{1.\arabic*\}},leftmargin=15mm]
\setlength\itemsep{0em}
\item a vacancy V inside the $(i,j)$ subgrid is picked;
\item said V is on one of the edges of the $(i,j)$ subgrid;
\item the cell picked to do the exchange is an A cell and is in one of the neighbouring $(i+1,j)$, $(i-1,j)$, $(i,j+1)$, $(i,j-1)$ subgrids.
\end{enumerate}
Similarly, to remove one A cell, it is necessary that at the $n$-th time step
\begin{enumerate}[label={\{2.\arabic*\}},leftmargin=15mm]
\setlength\itemsep{0em}
\item a vacancy V in one of the neighbouring $(i+1,j)$, $(i-1,j)$, $(i,j+1)$, $(i,j-1)$ subgrids is picked;
\item said V is on one of the edges of the subgrid it is on (the one neighbouring the $(i,j)$ subgrid);
\item the cell picked to do the exchange is an A cell and is in the $(i,j)$ subgrid.
\end{enumerate}

The probabilites of the events mentioned above are obtained via standard probability theory under the assumption that A and V are uniformly distributed in the subdomain,
%Let us define the following probabilities of the events mentioned above under the assumption that A and V are uniformly distributed in the subdomain,
\begin{align}
P(\{1.1\}) &= \frac{N \bar{V}\subij}{N^2 \bar{V}}, \qquad P(\{1.2\}) = 4\frac{\sqrt{N}}{N}, \\
\label{A-exchange-prob}
P(\{1.3\}) &= \frac{1}{4}\left(\bar{A}\subipj + \bar{A}\subimj + \bar{A}\subijp + \bar{A}\subijm \right),\end{align}
and
\begin{align}
P(\{2.1\}) &= \frac{N}{N^2 \bar{V}}\left( \bar{V}\subimj + \bar{V}\subipj + \bar{V}\subijm + \bar{V}\subijp \right), \\
P(\{2.2\}) &= \frac{\sqrt{N}}{N}, \qquad P(\{2.3\}) = \bar{A}\subij.
\end{align}

Now, the fraction of A cells at the next time step $\bar{A}\subij\np$ is just the fraction of A cells at the current time step $\bar{A}\subij^n$ plus the probability of adding an A cell into the subgrid, P(\{1.1\}) $\times$ P(\{1.2\}) $\times$ P(\{1.3\}), minus the probability of removing an A cell into the subgrid, P(\{2.1\}) $\times$ P(\{2.2\}) $\times$ P(\{2.3\}),
\begin{align}
\label{A-change}
\begin{split}
\bar{A}\subij\np = &\bar{A}\subij^n + \frac{1}{N} \left( \frac{N \bar{V}\subij^n}{N^2 \bar{V}} \right) \left( 4\frac{\sqrt{N}}{N} \right) \frac{1}{4}\left(\bar{A}\subipj^n + \bar{A}\subimj^n + \bar{A}\subijp^n + \bar{A}\subijm^n \right)
\\
&-\frac{1}{N} \left( \frac{N}{N^2 \bar{V}}\left( \bar{V}\subimj^n + \bar{V}\subipj^n + \bar{V}\subijm^n + \bar{V}\subijp^n \right) \right) \left( \frac{\sqrt{N}}{N} \right) \bar{A}\subij^n.
\end{split}
\end{align}

\subsection{V cells}

As in the previous case, there are two things can affect the amount of V cells in the $(i,j)$ subgrid: either an A or B cell in the $(i,j)$ subgrid is able to exchange places with a V cell in one of the neighbouring subgrids (that adds a V cell), or a vacancy from the $(i,j)$ subgrid is able to exchange places with an A or B cell in one of the neighbouring subgrids (that removes a V cell). To add one vacancy, it is necessary that at the $n$-th time step
\begin{enumerate}[label={\{3.\arabic*\}},leftmargin=15mm]
\setlength\itemsep{0em}
\item a vacancy V in one of the neighbouring $(i+1,j)$, $(i-1,j)$, $(i,j+1)$, $(i,j-1)$ subgrids is picked;
\item said V is on one of the edges of the subgrid it is on (the one neighbouring the $(i,j)$ subgrid);
\item the cell picked to do the exchange is on the $(i,j)$ subgrid and an exchange actually occurs. Recall that if the exchange is with an A cell the probability of movement is $1$ whereas if the exchange is with a B cell said probability is $1/\Gamma$.
\end{enumerate}
Similarly, to remove a vacancy, it is necessary that at the $n$-th time step
\begin{enumerate}[label={\{4.\arabic*\}},leftmargin=15mm]
\setlength\itemsep{0em}
\item a vacancy V inside the $(i,j)$ subgrid is picked;
\item said V is on one of the edges of the $(i,j)$ subgrid;
\item the cell picked to do the exchange is on one of the neighbouring $(i+1,j)$, $(i-1,j)$, $(i,j+1)$, $(i,j-1)$ subgrids and an exchange actually occurs. 
\end{enumerate}
It is assumed that the vacancy concentration is low enough so that the probability of picking a vacancy to do the exchange is negligible. Consequently we only define the probabilities 
\begin{align} 
P(\{3.3\}) &=  1-\bar{B}_{i,j} \frac{\Gamma - 1}{\Gamma},\\
P(\{4.3\}) &= 1-\frac{1}{4} (\bar{B}\subipj + \bar{B}\subimj + \bar{B}\subijp + \bar{B}\subijp) \frac{\Gamma - 1}{\Gamma}.
\end{align}
This allows us to write
\begin{align}
\label{V-change}
\begin{split}
\bar{V}\subij\np = &\bar{V}\subij + \frac{1}{N} \left( \frac{N}{N^2 \bar{V}}\left( \bar{V}\subimj^n + \bar{V}\subipj^n + \bar{V}\subijm^n + \bar{V}\subijp^n \right) \right) \left( \frac{\sqrt{N}}{N} \right) \left( 1 - \bar{B}_{i,j}^n \frac{\Gamma - 1}{\Gamma} \right) \\
&-\frac{1}{N} \left( \frac{N V\subij^n}{N^2 \bar{V}} \right) \left( 4 \frac{\sqrt{N}}{N}\right) \left( 1 - \frac{1}{4} (\bar{B}\subipj^n + \bar{B}\subimj^n + \bar{B}\subijp^n + \bar{B}\subijp^n)\frac{\Gamma - 1}{\Gamma} \right).
\end{split}
\end{align}

\subsection{Limit $N \rightarrow \infty$}
\label{sec-limit}

The objective of the present section is to determine whether in the limit $N \rightarrow \infty$ the CA model of the previous section reduces to the diffusion model in equations \eqref{gov-eq-1-mathmodel}-\eqref{gov-eq-2-mathmodel}.
%The objective of the present section is to take the continuum limit of the cellular automaton model we presented in this paper. %to gain insight in the physical problem described in \cite{Ribera2017}, summarised in \S \ref{sec-model-general}. 
Rearranging equation \eqref{A-change} leads to
\begin{align}
\label{A-change-limit}
\begin{split}
\bar{A}\subij\np = &\bar{A}\subij^n + \frac{\sqrt{N}}{N^2 \bar{V}} \left( \frac{\bar{V}\subij^n \left( \bar{A}\subimj^n + \bar{A}\subipj^n + \bar{A}\subijm^n + \bar{A}\subijp^n \right)}{N} \right. \\
&\left.-\frac{\left( \bar{V}\subimj^n + \bar{V}\subipj^n + \bar{V}\subijm^n + \bar{V}\subijp^n \right) \bar{A}\subij^n}{N} \right).
\end{split}
\end{align}
We may equate this to a standard finite difference form by choosing $\sqrt{N} = \Delta x$ and $\Delta \hat{t} = \frac{1}{N^{3/2} \bar{V}}$,
\begin{align}
\label{A-change-limit-2}
\begin{split}
\frac{\bar{A}\subij\np - \bar{A}\subij^n}{\Delta \hat{t}} = &\left( \frac{\bar{V}\subij^n \left( \bar{A}\subimj^n + \bar{A}\subipj^n + \bar{A}\subijm^n + \bar{A}\subijp^n \right)}{\Delta x^2} \right. \\
&\left.-\frac{\left( \bar{V}\subimj^n + \bar{V}\subipj^n + \bar{V}\subijm^n + \bar{V}\subijp^n \right) \bar{A}\subij^n}{\Delta x^2} \right).
\end{split}
\end{align}
Note that $\Delta \hat{t}$ is not a time, rather we are just conforming to finite difference notation.

Similarly, equation \eqref{V-change} may be expressed as 
\begin{align}
\label{V-change-limit-2}
\begin{split}
\frac{\bar{V}\subij\np - \bar{V}\subij^n}{\Delta \hat{t}} = &\frac{\left( \bar{V}\subimj^n + \bar{V}\subipj^n + \bar{V}\subijm^n + \bar{V}\subijp^n \right)   \left( 1 - \bar{B}_{i,j}^n \frac{\Gamma - 1}{\Gamma} \right)}{\Delta x^2} \\
&-  \frac{\bar{V}\subij^n \left( 4 - (\bar{B}\subipj^n + \bar{B}\subimj^n + \bar{B}\subijp^n + \bar{B}\subijp^n)\frac{\Gamma - 1}{\Gamma} \right)}{\Delta x^2}.
\end{split}
\end{align}

Taking the limit $N \rightarrow \infty$ in equations \eqref{A-change-limit-2} and \eqref{V-change-limit-2}, changing the notation to $A\subij \equiv X_A$, $B\subij \equiv X_B$, $V\subij \equiv X_V$, and substituting $X_B = 1 - X_A - X_V$ gives
\begin{align}
\label{A-change-limit-4}
\frac{\partial X_A}{\partial \hat{t}} &= X_V \nabla^2 X_A - X_A \nabla^2 X_V,\\
\label{V-change-limit-4}
\Gamma \frac{\partial X_V}{\partial \hat{t}} &=  -(\Gamma - 1) X_V \nabla^2 X_A + \left( 1 + (\Gamma - 1)X_A \right)\nabla^2 X_V.
\end{align}

Finally, setting $t = \hat{t}/\Gamma$, we recover equations \eqref{gov-eq-1-mathmodel}-\eqref{gov-eq-2-mathmodel} and the CA model does indeed reduce to the continuous diffusion model.

\section{Results}

In this section we present the results of the CA model. For the simulations we pick a square grid $N \times N$, where $N=200$. At the first time step, the first 100 columns are A cells, and the remaining 100 columns are B cells. Then we randomly distribute 2000 vacancies (equivalent to 5\% of the total number of cells) throughout the whole grid. A simulation is then run for $1.85 \times 10^9$ steps which is sufficient to allow for significant change in the distribution of cells.

\begin{figure}[!htb]
\begin{center}
\begin{subfigure}[b]{0.49\textwidth}
\begin{center}
\includegraphics[width=0.78\textwidth]{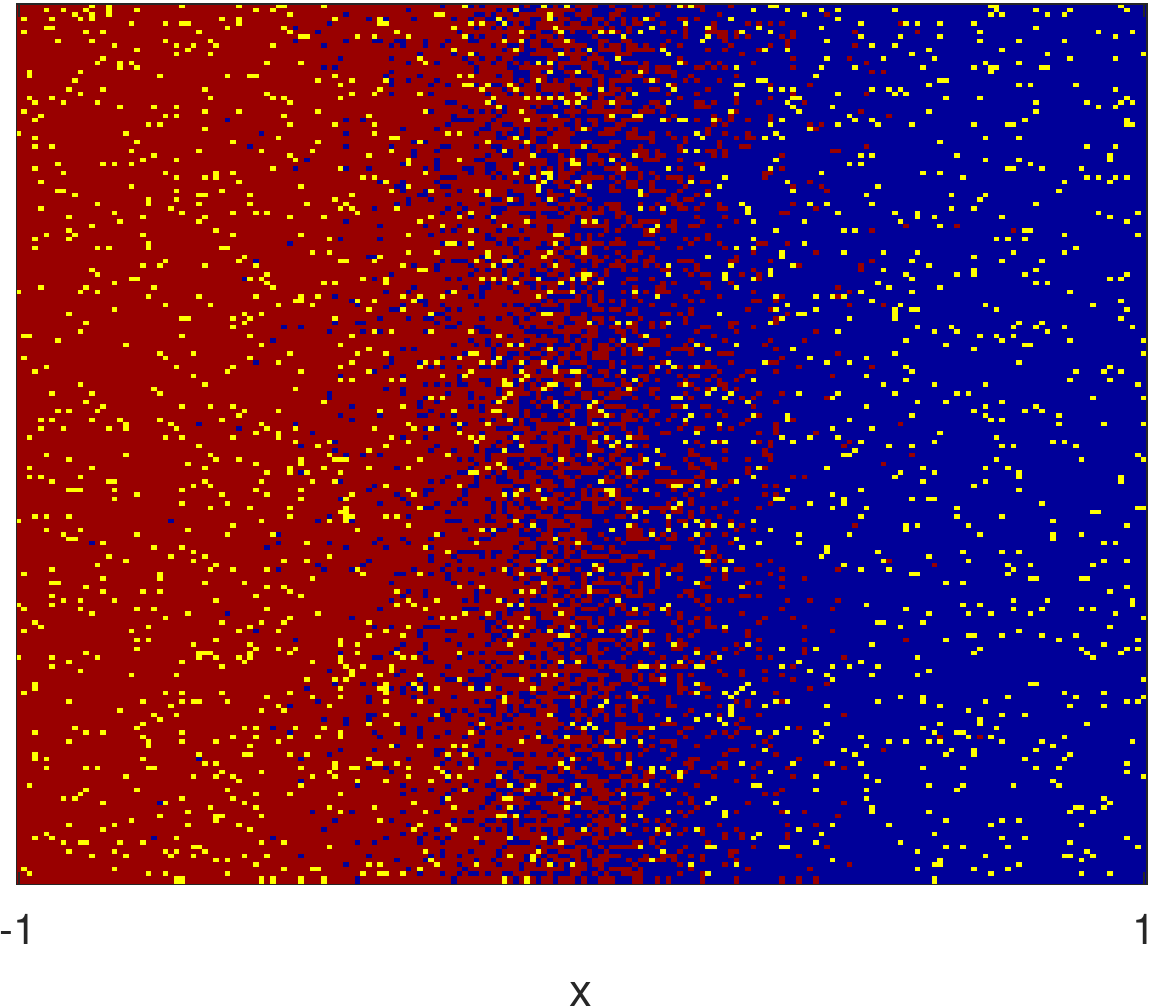}
\end{center}
\vskip-4mm
\caption{$\Gamma = 1.5$, $t = t_1$.}
\end{subfigure}
\begin{subfigure}[b]{0.49\textwidth}
\begin{center}
\includegraphics[width=0.78\textwidth]{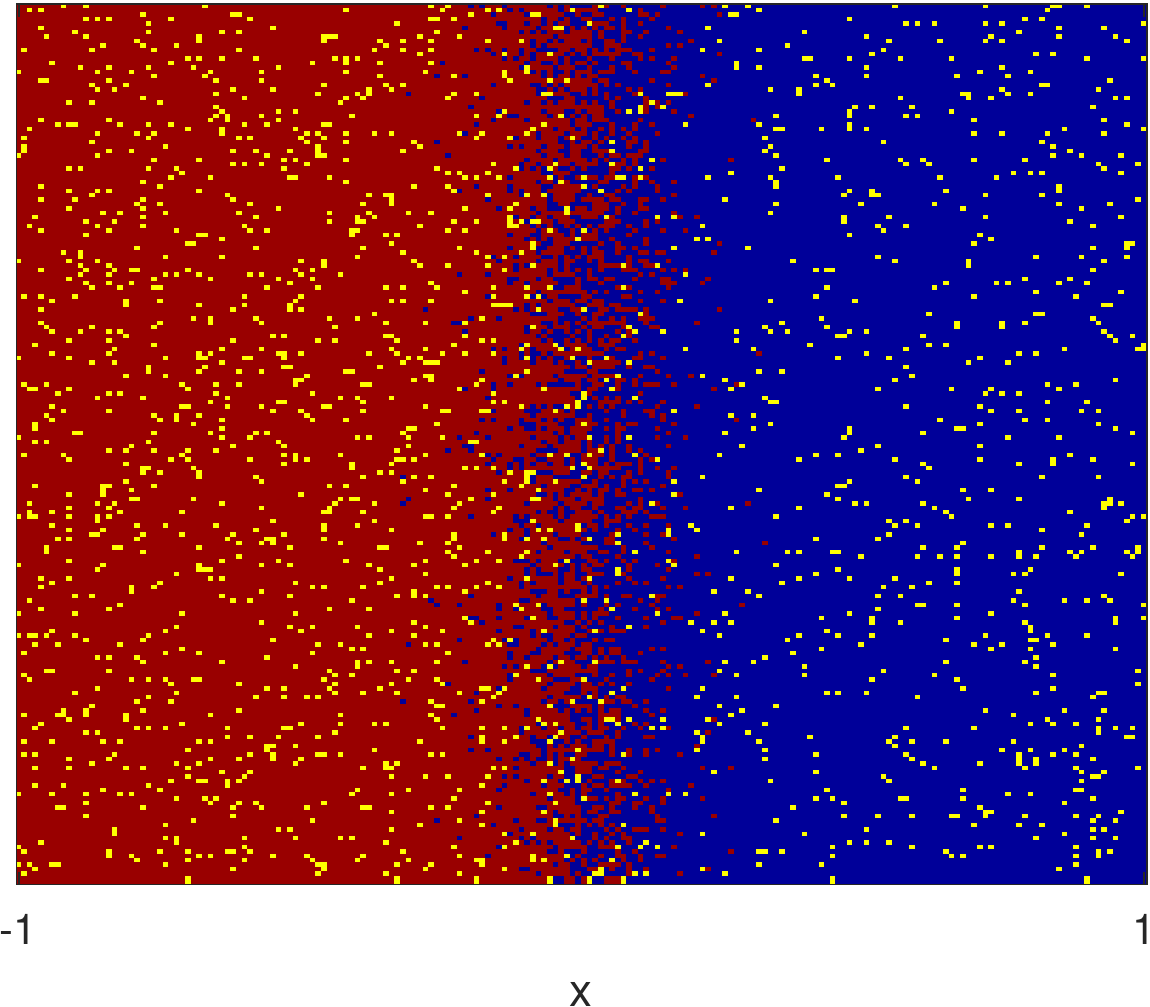}
\end{center}
\vskip-4mm
\caption{$\Gamma = 10$, $t = t_1$.}
\end{subfigure}

\vskip5mm

\begin{subfigure}[b]{0.49\textwidth}
\begin{center}
\includegraphics[width=0.78\textwidth]{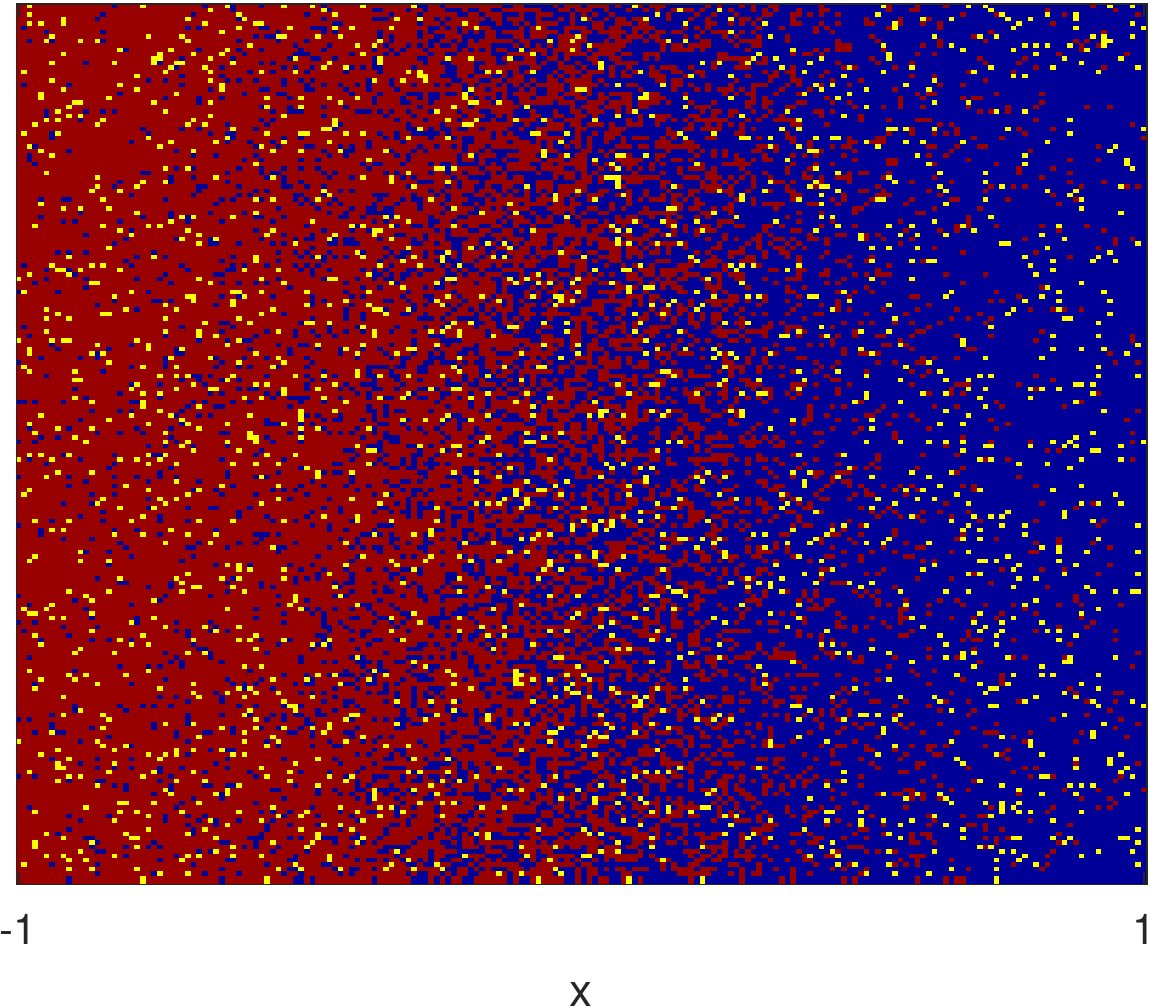}
\end{center}
\vskip-4mm
\caption{$\Gamma = 1.5$, $t = t_2$.}
\end{subfigure}
\begin{subfigure}[b]{0.49\textwidth}
\begin{center}
\includegraphics[width=0.78\textwidth]{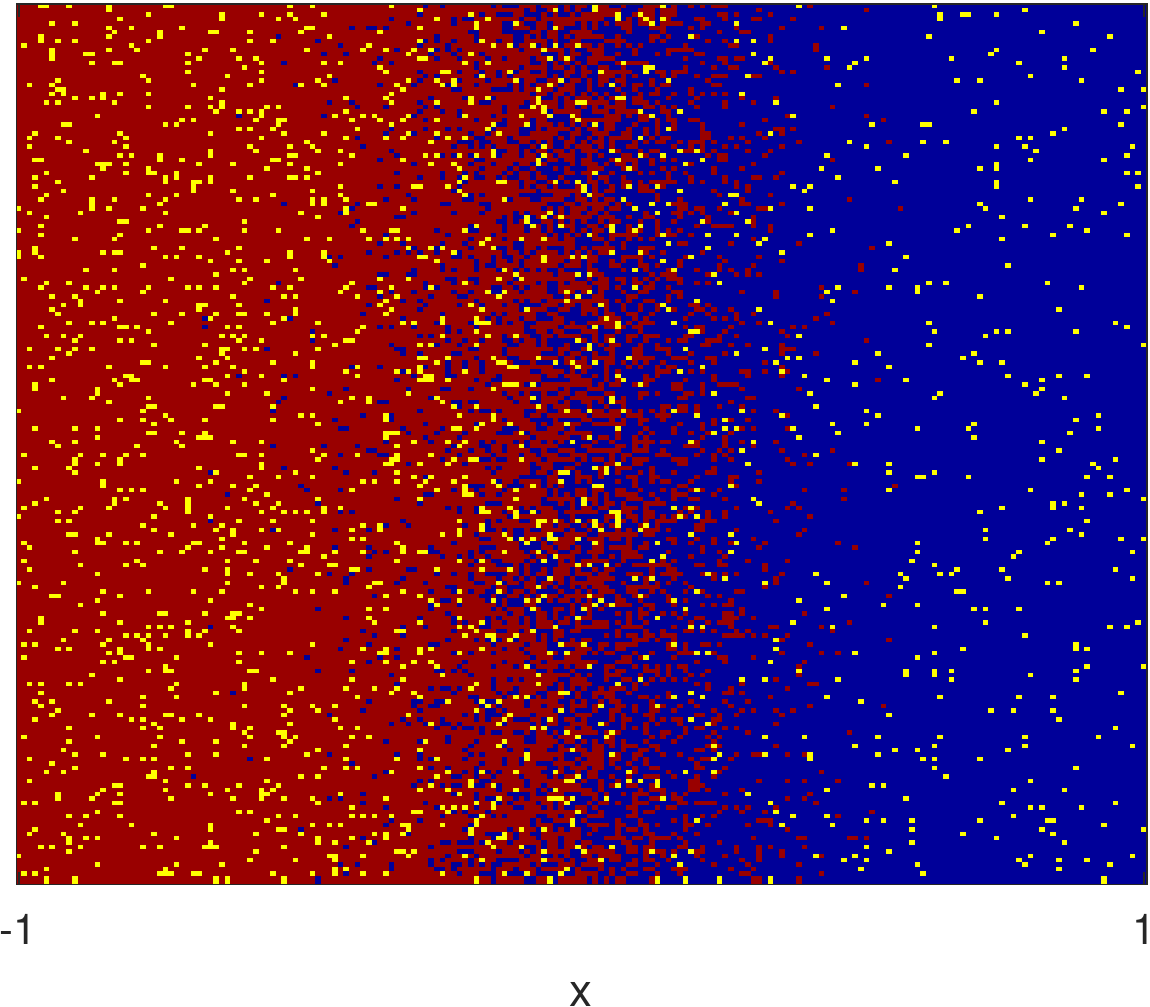}
\end{center}
\vskip-4mm
\caption{$\Gamma = 10$, $t = t_2$.}
\end{subfigure}

\vskip5mm

\begin{subfigure}[b]{0.49\textwidth}
\begin{center}
\includegraphics[width=0.78\textwidth]{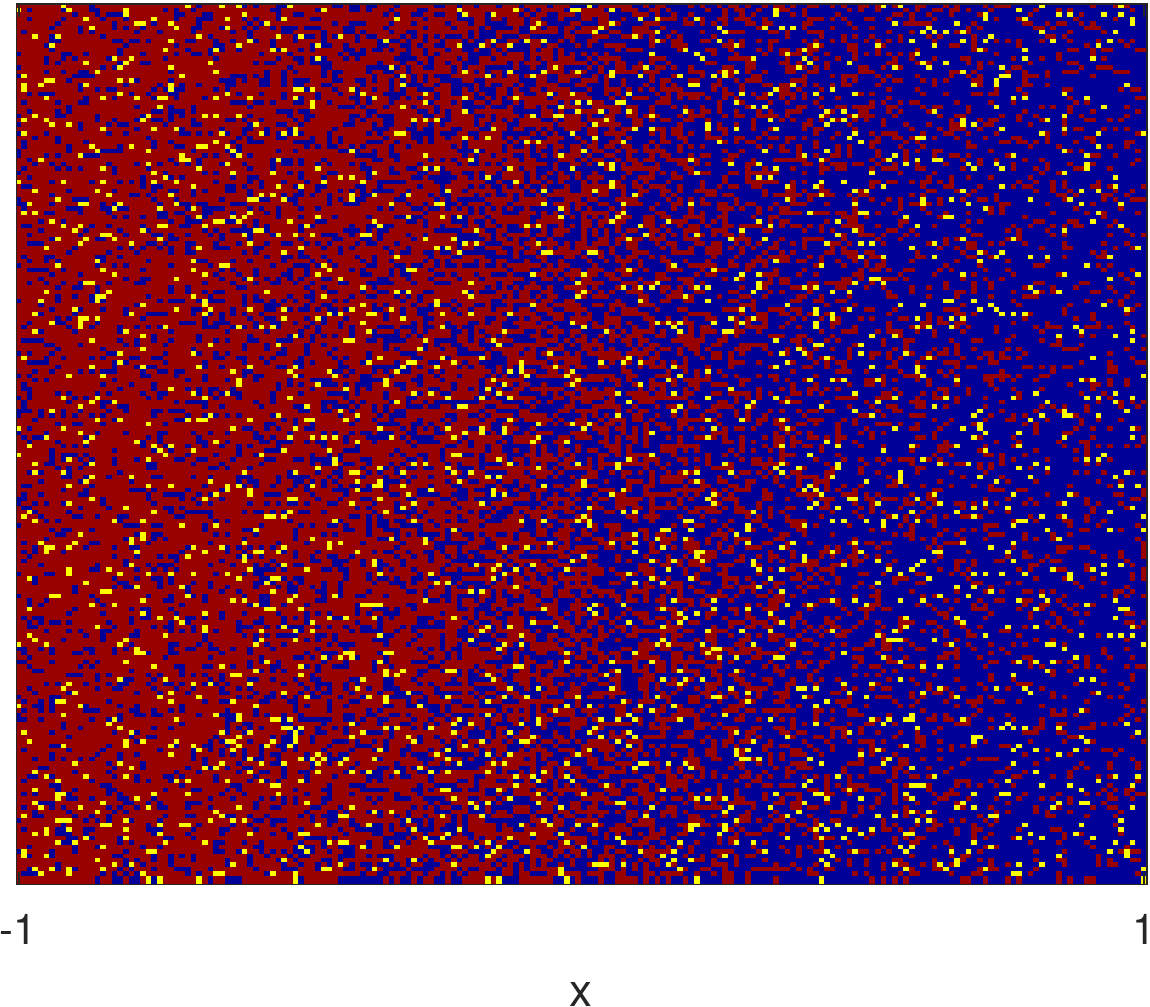}
\end{center}
\vskip-4mm
\caption{$\Gamma = 1.5$, $t = t_3$.}
\end{subfigure}
\begin{subfigure}[b]{0.49\textwidth}
\begin{center}
\includegraphics[width=0.78\textwidth]{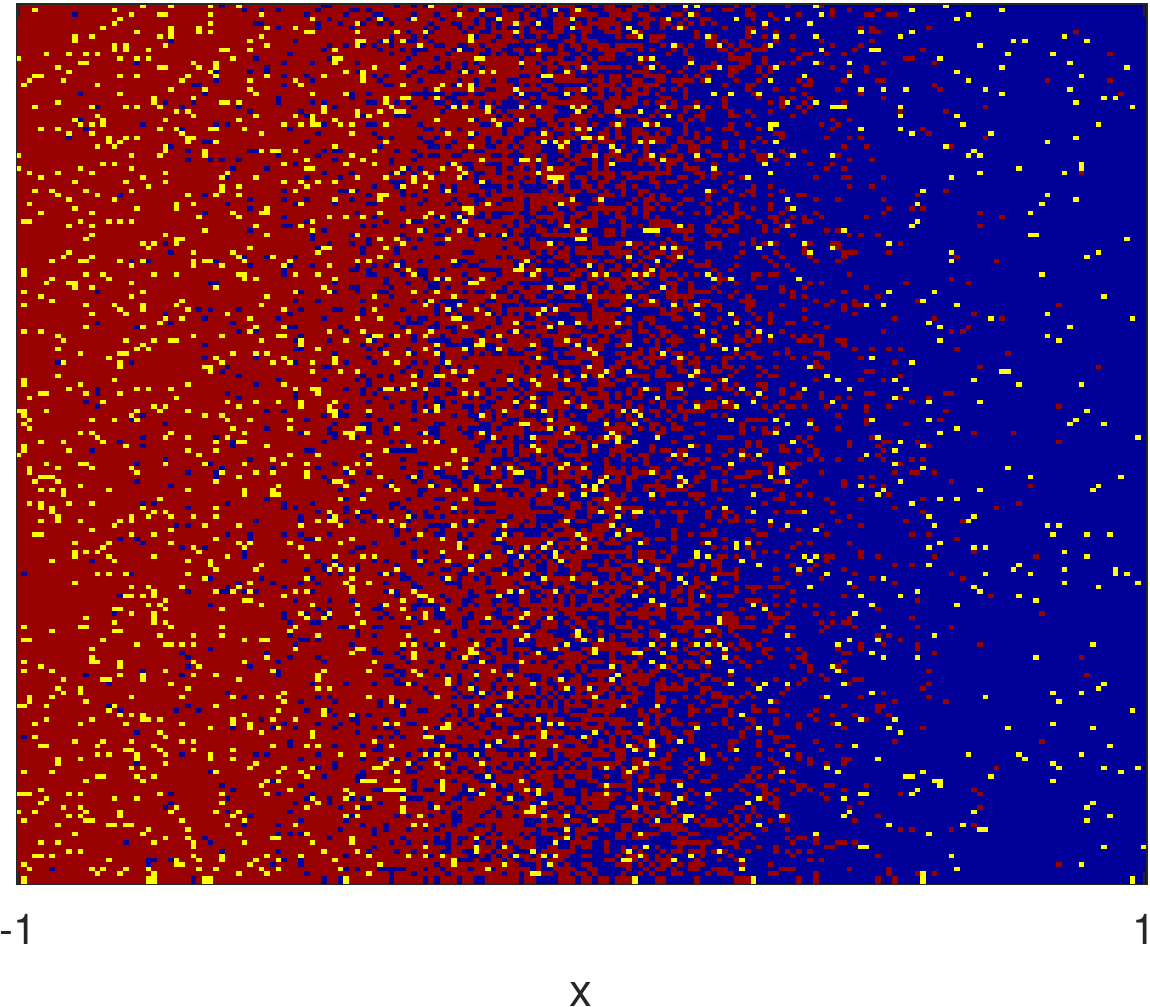}
\end{center}
\vskip-4mm
\caption{$\Gamma = 10$, $t = t_3$.}
\end{subfigure}

\caption{Resulting grid $200 \times 200$ for different times obtained with the simulation of the CA model. Red denotes A atom cells. Blue denotes B atom cells. Yellow denotes vacancy cells. $t_1 < t_2 < t_3$.}
\label{figure-CA-mesh-different-times}
\end{center}
\end{figure}

Figures \ref{figure-CA-mesh-different-times}(a), \ref{figure-CA-mesh-different-times}(c), \ref{figure-CA-mesh-different-times}(e) show the distribution of material and vacancies when $\Gamma = 1.5$, that is, the diffusion rates between A and B are similar. Throughout the process vacancies are well distributed in the domain and, by the time $t=t_3$, the system appears close to equilibrium. Figures \ref{figure-CA-mesh-different-times}(b), \ref{figure-CA-mesh-different-times}(d), \ref{figure-CA-mesh-different-times}(f) show the corresponding evolution when A diffuses much faster than B. In Figure \ref{figure-CA-mesh-different-times}(b) we see a greater motion of A to the right than the one observed in Figure \ref{figure-CA-mesh-different-times}(a). However this also means that vacancies accumulate on the left. In Figure \ref{figure-CA-mesh-different-times}(d) it is clear that the vacancy concentration on the right is low, which acts to slow down the diffusion. This is clear from the final figure, Figure \ref{figure-CA-mesh-different-times}(f), which is far from equilibrium. This may seem counter-intuitive; A diffuses much faster here than in Figure \ref{figure-CA-mesh-different-times}(e) but it clearly ends up moving slower. This is a result of the initial rapid movement of A, bringing a high proportion of vacancies to the left and so hindering further movement. A similar result was noticed in the continuum model of \cite{Ribera2017}, where a very fast diffuser ends up redistributing more slowly.

\begin{figure}[!htb]
\begin{center}
\begin{subfigure}[b]{0.49\textwidth}
\includegraphics[width=\textwidth=1]{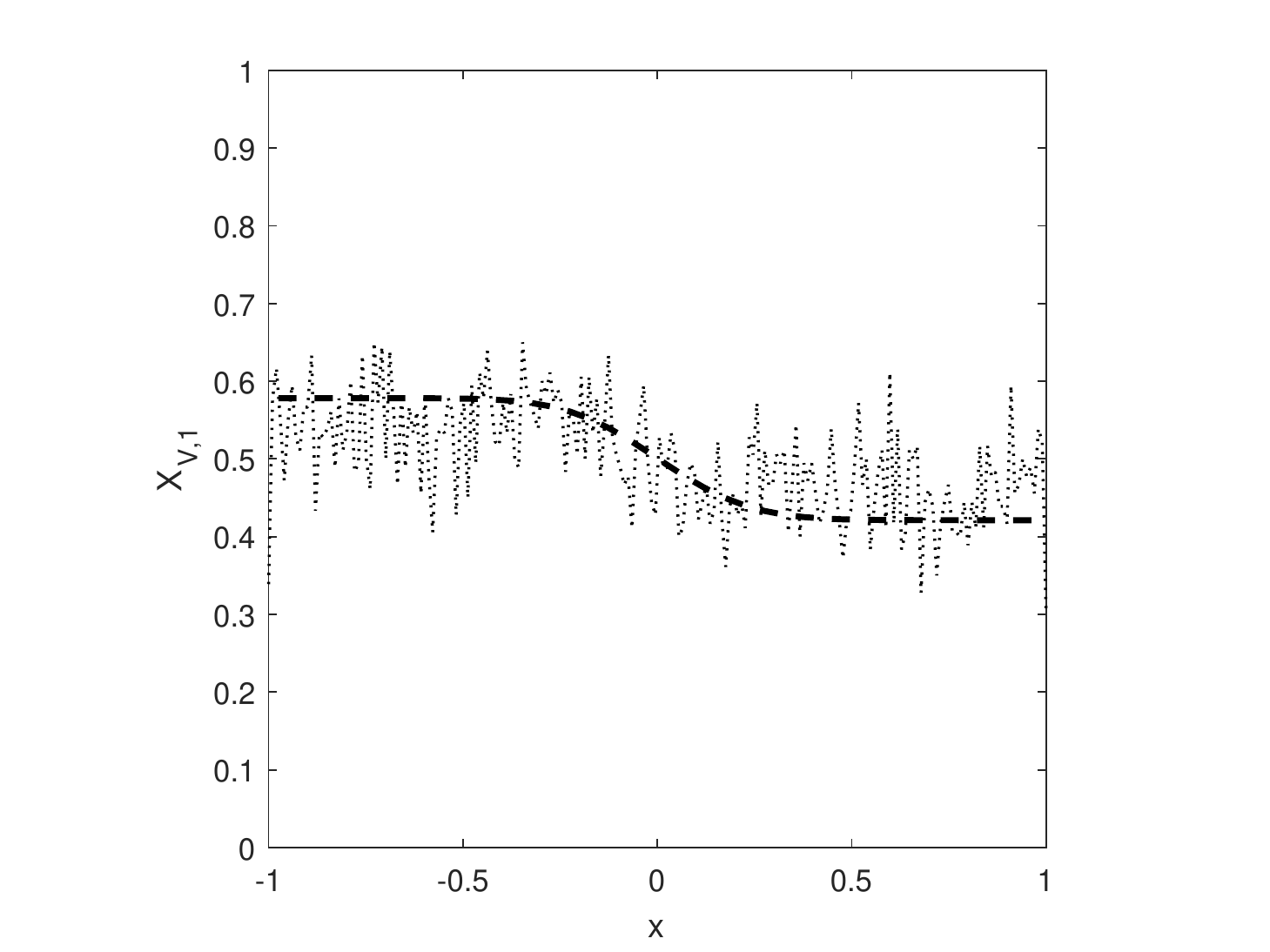}
\caption{$\Gamma = 1.5$, $t = 0.0378$.}
\end{subfigure}
\begin{subfigure}[b]{0.49\textwidth}
\includegraphics[width=\textwidth=1]{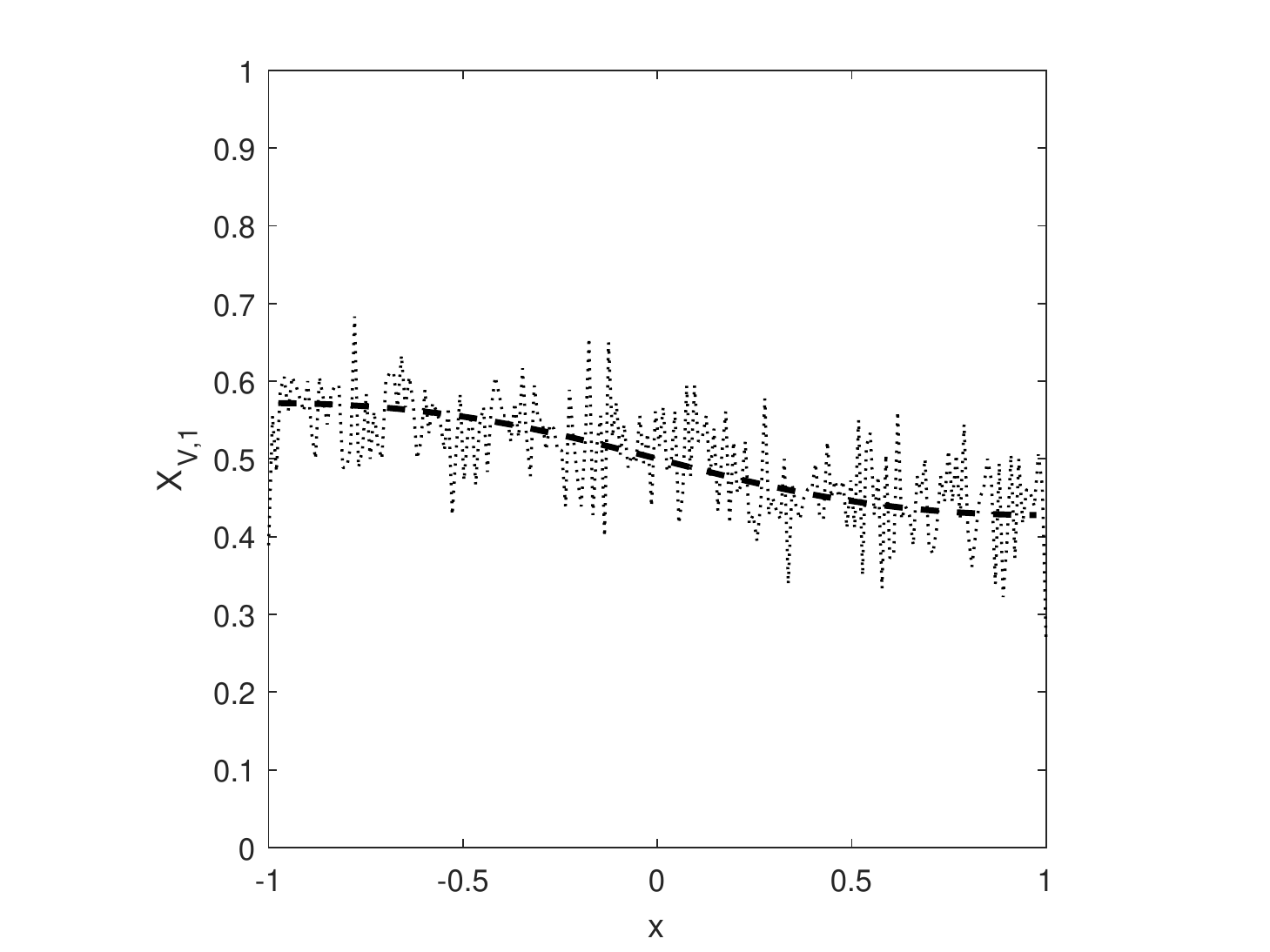}
\caption{$\Gamma = 1.5$, $t = 0.2519$.}
\end{subfigure}
\begin{subfigure}[b]{0.49\textwidth}
\includegraphics[width=\textwidth=1]{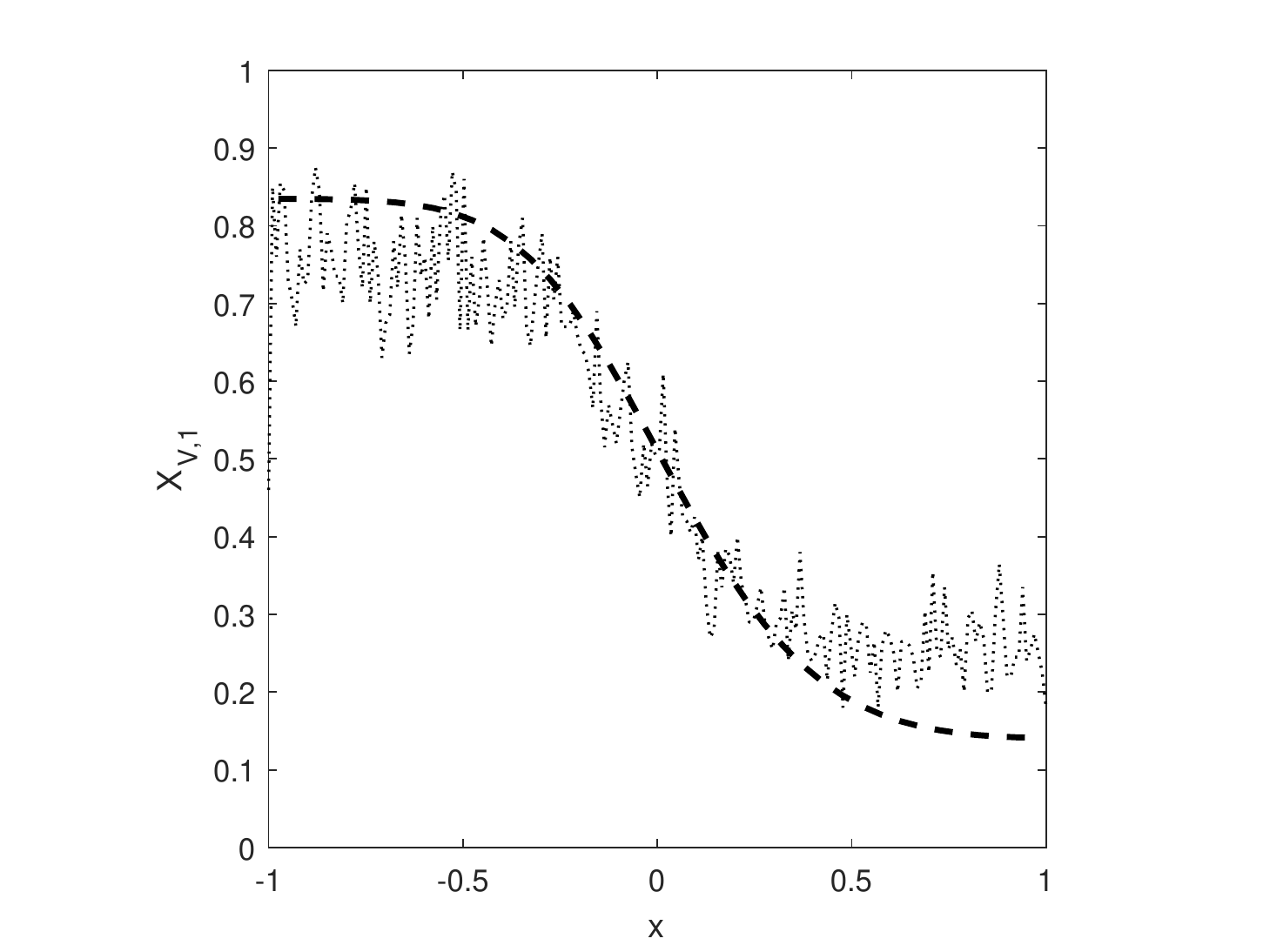}
\caption{$\Gamma = 10$, $t = 0.0574$.}
\end{subfigure}
\begin{subfigure}[b]{0.49\textwidth}
\includegraphics[width=\textwidth=1]{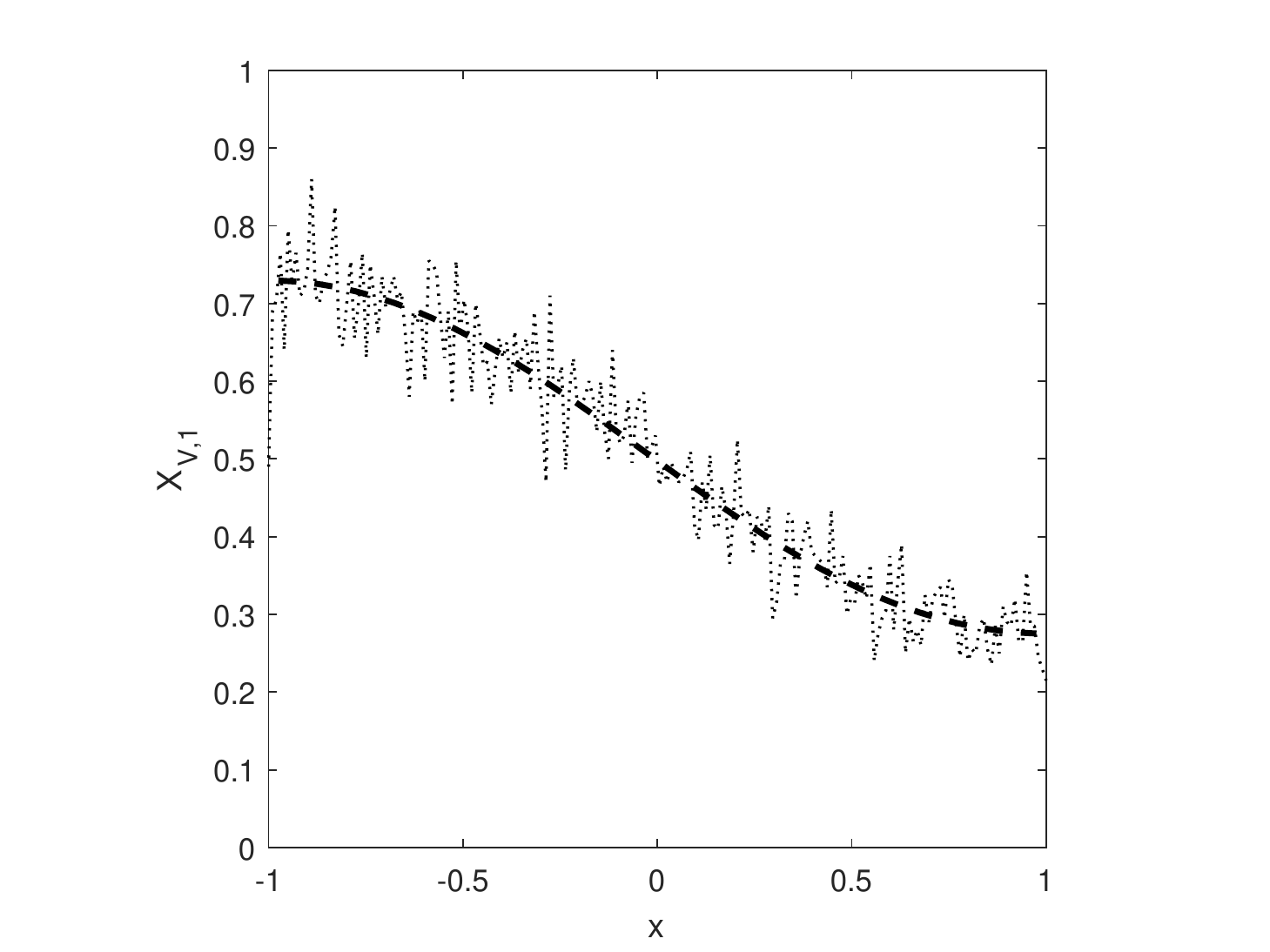}
\caption{$\Gamma = 10$, $t = 0.3263$.}
\end{subfigure}
\caption{Dotted line is obtained by joining the discrete normalised vacancy average concentration $V_j^n$ obtained via the CA model (equation \eqref{average-V}). Solid line represents the numerical solution of the continuum model described by equations \eqref{gov-eq-1-mathmodel}-\eqref{gov-eq-2-mathmodel}.}
\label{figure-CA-vs-FullNumerics}
\end{center}
\end{figure}

In Figure \ref{figure-CA-vs-FullNumerics} we compare the concentration of vacancies given by the CA model and that given by the continuum model in equations \eqref{gov-eq-1-mathmodel}-\eqref{gov-eq-2-mathmodel} at different times for two values of ratio between the diffusion rates; $\Gamma = 1.5$ (Figure \ref{figure-CA-vs-FullNumerics}(a) and \ref{figure-CA-vs-FullNumerics}(b)) and $\Gamma = 10$ (Figure \ref{figure-CA-vs-FullNumerics}(c) and \ref{figure-CA-vs-FullNumerics}(d)). The numerical solution of \eqref{gov-eq-1-mathmodel}-\eqref{gov-eq-2-mathmodel} is standard and defined in \cite{Ribera2017}. The CA results come from an average of simulations. To achieve this we define
\begin{equation}
\label{average-V}
V_j^n = \frac{\bar{V}}{I} \sum_{i=1}^I \left\{N_V^i \right\}^n_j
\end{equation}
where $\left\{N_V^i \right\}^n_j$ is the number of vacancies at the $j$-th column on the $n$-th time step, the superscript $i$ is used to distinguish different simulations, and $I$ is the total number of CA simulations. The variable $V_j^n$ denotes the average concentration of vacancies at the $j$-th column on the $n$-th time step. The results shown in Figure \ref{figure-CA-vs-FullNumerics} correspond to $I=10$. 

To be able to compare the variable $V_j^n$ and the numerical solution $X_V$ of \eqref{gov-eq-1-mathmodel}-\eqref{gov-eq-2-mathmodel} we need to find a correspondence between the discrete time in the CA simulation and the continuous time in the PDE system. Let $n$ be the time step that needs to be transformed to a continuum time $t_n$. Then
\begin{equation}
t_n = n \frac{\Delta \hat{t}}{\Gamma} t_s^{-1},
\end{equation}
where $t_s$ is the time scale defined in \cite{Ribera2017}. It is defined as $t_s =  \lambda a^2 \Gamma_B/(\epsilon l^2)$,
where $\epsilon = 2 X_{V,\text{ini}}$, $\lambda$ is a geometric factor, $a$ is the lattice constant and $\Gamma_B$ is the hop frequency of species B. Finally, $l$ is the length of the one-dimensional bar and it is found via the lattice site density $\rho$ and the number of cells in our CA grid,
\begin{equation}
l = \left(\frac{N^3}{\rho}\right)^{1/3},
\end{equation}
which gives $l=51$ nm for the case presented in Figure \ref{figure-CA-vs-FullNumerics}. Using the parameters values shown in \cite[Table 1]{Ribera2017}, we find that $t_s \approx 1.05 \times 10^{8}$. There is only one issue remaining to be treated before being able to compare the two solutions. When solving the continuum model, the initial condition for vacancies is given by \cite{Ribera2017}
\begin{equation}
X_{V,1}(x,0) = \frac{1+(\Gamma-1)X_A(x,0)}{2+(\Gamma-1)X_{A,\text{ini}}},
\end{equation}
where $X_{V,1} = X_V/\epsilon$. We do this rescaling to be able to keep track of the evolution of vacancies since this number is usually very small compared to the concentration of species A and B; numerically, this may cause problems. The initial condition for the CA model corresponds to $X_{V,1}(x,0) \approx 0.5$. This merely means that if $n_0$ is the initial time step in the CA model, the actual $n_{t_0}$ that corresponds to the initial time $t_0$ is $n_{t_0} = r n_0$, where $r > 1$ (see Figure \ref{sketch-conversion}). To find this rescaling factor $r$ we minimise the least-squares error between the continuum model and the discrete set of data $V_j^{n_f}$ at $n_f=1.85\times 10^9$. For $\Gamma = 10$, we find that $r = 26.171$; for $\Gamma = 1.5$, $r = 21.528$. %We use $r_{\hat{t}} = 0.31$ for both cases.
The comparison of $V_j^{n}$, for $j=1,\dots,\,N$ and $X_{V,1}(t,x)$, with $x\in [-1,\,1]$, is now well defined.
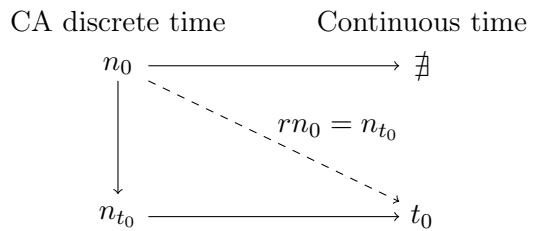
\begin{wrapfigure}{R}{7.5cm}
\centering
\begin{tikzpicture}
\node at (0,0.6) {CA discrete time};
\node at (4,0.6) {Continuous time};
\node at (0,0) {$n_0$};
\draw[arrows=->] (0.4,0) -- (3.7,0);
\draw[arrows=->] (0,-0.2) -- (0,-1.7);
\node at (4,0) {$\nexists$};

\node at (0,-2) {$n_{t_0}$};
\draw[arrows=->] (0.4,-2) -- (3.7,-2);
\node at (4,-2) {$t_0$};

\draw[arrows=->,dashed] (0.4,-0.2) -- (3.7,-1.8);

\node at (2.9,-0.85) {$r n_0 = n_{t_0}$};

\end{tikzpicture}
\caption{Time conversion sketch.}
\label{sketch-conversion}
\label{sketch-conversion}
\end{wrapfigure}

%\begin{figure}[!htb]
%\centering
%\begin{tikzpicture}
%\node at (0,0.6) {CA discrete time};
%\node at (4,0.6) {Continuous time};
%\node at (0,0) {$n_0$};
%\draw[arrows=->] (0.4,0) -- (3.7,0);
%\draw[arrows=->] (0,-0.2) -- (0,-1.7);
%\node at (4,0) {$\nexists$};
%
%\node at (0,-2) {$n_{t_0}$};
%\draw[arrows=->] (0.4,-2) -- (3.7,-2);
%\node at (4,-2) {$t_0$};
%
%\draw[arrows=->,dashed] (0.4,-0.2) -- (3.7,-1.8);
%
%\node at (2.9,-0.85) {$r n_0 = n_{t_0}$};
%
%\end{tikzpicture}
%\caption{Time conversion sketch.}
%\label{sketch-conversion}
%\end{figure}

In Figure \ref{figure-CA-vs-FullNumerics} we compare results for the vacancy concentration from the continuum model of equations \eqref{gov-eq-1-mathmodel}-\eqref{gov-eq-2-mathmodel} and the average result of 10 simulations via equation \eqref{average-V}. First, we note what was observed in the previous figures, when $\Gamma = 1.5$ the vacancy concentration is relatively constant. When $\Gamma = 10$ vacancies concentrate on the left, thus slowing the movement of the fast diffuser. All figures show good agreement, even when $\Gamma = 1.5$. The most noticeable discrepancies occur at small times, near the ends $x=\pm 1$, where the continuum model indicates greater movement from the initial condition ($X_V = 0.5$). This is not surprising, continuum diffusion models typically allow for motion throughout the domain even though in reality extreme points may not be feeling any effect. Hence we expect that for small times the CA model is more realistic, and the agreement improves with time.

\section{Conclusions}

The goal of this paper was to develop a cellular automaton model to describe binary diffusion in solids. To do this we chose to employ an asynchronous model and used basic rule definitions to update the cell states, based on the physics that drive the process.

In the limit of large number of cells we showed that this CA model reduces to a particular form of the continuum model developed in \cite{Ribera2017}. This was verified in the Results section by comparing the CA and the continuum model. This opens the possibility of designing various scenarios in a very simple fashion and then just take the limit to obtain the continuum model to do a more accurate analysis.

An interesting result to come out of this work is that when the system has a very fast diffuser it can lead, overall, to a slower diffusion process. This occurs because the initial fast diffusion acts to move nearly all vacancies to one side, so restricting further vacancy exchange and so movement. Also, at small times the nature of the continuum model permitted movement throughout the domain when in practice this may not occur. The CA model showed less movement near the extreme points at small times. This seems more physically realistic, hence in this case the CA model may be preferable. 

Obviously the CA model becomes increasingly cumbersome as the number of cells increases. For sufficiently large numbers a continuum model is clearly preferable. However, when the number of cells is small, for example when modelling nanoscale diffusion, CA models provide a powerful tool which may be more accurate than the continuum models.

%Furthermore, we have showed that modelling binary diffusion in solids via a cellular automaton approach, using a very simple rule, leads to sensible simplified governing equations presented in \cite{Ribera2017}. Indeed, in the Results section we have showed that the cellular automaton approach is able to qualitatively approximate the results of the equivalent continuum model. This opens the possibility of designing various scenarios in a very simple fashion and then just take the limit to obtain the continuum model to do a more accurate analysis. This type of technique for modelling binary diffusion could be a powerful tool to obtain mathematical models of different diffusion processes, such as the one that allows the creation of hollow structures at the nanoscale.

%%%%%%%%%%%%%%%%%%%%

\section*{Acknowledgements}

HR and TM acknowledge that the research leading to these results has received funding from ``la Caixa'' Foundation and has been partially funded by the CERCA Programme of the Generalitat de Catalunya. TM acknowledges Ministerio de Ciencia e Innovación grant MTM2014-5621. BW acknowledges an NSERC Canada research grant.

%%%%%%%%%%%%%%%%%%%%

\clearpage

%\bibliographystyle{abbrv}
%%\nocite{*}
%\inputencoding{utf8}
%\bibliography{../Kirkendall-effect,../cellular-automata}

\end{document}